\documentclass[conference]{IEEEtran}
\IEEEoverridecommandlockouts
\usepackage{cite}
\usepackage{amsmath,amssymb,amsfonts}
\usepackage[noend]{algpseudocode}
\usepackage{algorithm}
\usepackage{graphicx}
\usepackage{textcomp}
\usepackage{xcolor}
\usepackage{booktabs}
\usepackage{url}
\usepackage{array}
\usepackage[caption=false, font=small, labelfont=sf, textfont=sf]{subfig}

\newcolumntype{x}[1]{>{\centering\arraybackslash\hspace{0pt}}p{#1}}
\def\BibTeX{{\rm B\kern-.05em{\sc i\kern-.025em b}\kern-.08em
    T\kern-.1667em\lower.7ex\hbox{E}\kern-.125emX}}
\begin{document}

\title{UniGPS: A Unified Programming Framework for Distributed Graph Processing}

\author{\IEEEauthorblockN{Zhaokang Wang\IEEEauthorrefmark{1}\IEEEauthorrefmark{2}, Junhong Li\IEEEauthorrefmark{1}\IEEEauthorrefmark{2}, Yifan Qi\IEEEauthorrefmark{1}\IEEEauthorrefmark{2}, Guanghui Zhu\IEEEauthorrefmark{1}\IEEEauthorrefmark{2}, Chunfeng Yuan\IEEEauthorrefmark{1}\IEEEauthorrefmark{2}, Yihua Huang\IEEEauthorrefmark{1}\IEEEauthorrefmark{2}}
\IEEEauthorblockA{\IEEEauthorrefmark{1}\textit{State Key Laboratory for Novel Software Technology, Nanjing University}, Nanjing, P. R. China}
\IEEEauthorblockA{\IEEEauthorrefmark{2}\textit{Department of Computer Science and Technology, Nanjing University}, Nanjing, P. R. China\\
\{wangzhaokang,mg1833036,mg20330046\}@smail.nju.edu.cn, \{zgh,cfyuan,yhuang\}@nju.edu.cn}
%\IEEEauthorblockA{\IEEEauthorrefmark{3}\textit{Jiangsu Hongcheng Big Data Technology and Application Institute Co., Ltd.}, Nanjing, P. R. China\\
%\{guochen,liyifan\}@jshcbd.com}
}

\maketitle

\begin{abstract}
The industry and academia have proposed many distributed graph processing systems.
However, the existing systems are not friendly enough for users like data analysts and algorithm engineers.
On the one hand, the programing models and interfaces differ a lot in the existing systems, leading to high learning costs and program migration costs.
On the other hand,  these graph processing systems are tightly bound to the underlying distributed computing platforms, requiring users to be familiar with distributed computing. 
To improve the usability of distributed graph processing, we propose a unified distributed graph programming framework UniGPS.
Firstly, we propose a unified cross-platform graph programming model VCProg for UniGPS.
VCProg hides details of distributed computing from users.
It is compatible with the popular graph programming models Pregel, GAS, and Push-Pull.
VCProg programs can be executed by compatible distributed graph processing systems without modification, reducing the learning overheads of users.
Secondly, UniGPS supports Python as the programming language.
We propose an interprocess-communication-based execution environment isolation mechanism to enable Java/C++-based systems to call user-defined methods written in Python.
The experimental results show that UniGPS enables users to process big graphs beyond the memory capacity of a single machine without sacrificing usability.
UniGPS shows near-linear data scalability and machine scalability.
\end{abstract}


%! TEX root=unigps-main.tex
%! TEX spellcheck=en_US
\section{Introduction}

\subsection{Background}
\label{sec_backgroud}

The graph is a useful data structure that can model relationships between multiple entities in the real world, such as social networks and transaction graphs in e-commerce platforms.
The industry and academia have proposed many single-machine graph processing systems \cite{graph_processing_framework_survey_2018} to facilitate graph analysis.
They provide users with friendly programming models and application programming interfaces (API).
However, the computing power and memory space of a single machine limit their performance.
As graph scales in real-world applications grow rapidly, their performance on big graphs becomes more and more unsatisfactory.

In order to process big graphs efficiently, many distributed graph processing systems are proposed \cite{graph_processing_framework_survey_2018}.
They simplify distributed graph processing by providing users with high-level graph programming models and APIs.
%
%The programming models hide some distributed computing details (like communication and fault tolerance) from users.
%
However, they are still not easy to use for users not familiar with distributed computing (like data analysts and algorithm engineers).
They have the following shortcomings in usability.
%, which hinders the popularization of distributed graph processing among casual users.

\begin{itemize}
	\item The programming interfaces of the existing systems lack the cross-platform feature.
	Each system has its unique programming model and APIs.
	The programs written for a certain system can only run on that system.
	When a new system with higher performance appears, users need manually migrate the existing programs to the new system, introducing additional learning costs and programming costs.

	\item The programming interfaces of the mainstream systems (like Giraph \cite{giraph} and GraphX \cite{graphx}) are tightly bound to the underlying distributed computing platforms.
	Users have to learn how to use distributed computing platforms before using the distributed graph processing systems, creating extra barriers.
	For example, Giraph requires users to be familiar with Hadoop's APIs such as Writable and FileInputFormat to write correct Giraph programs.

	\item Few mainstream systems support Python as the programming language.
	The mainstream systems (Giraph, GraphX, and Gemini) only support compiled languages (Java, Scala, and C++), but the data analysts or algorithm engineers prefer using Python \cite{top_programming_language_survey_ieee}\cite{top_programming_language_survey_kdnuggets}.
	Python, along with the interactive development environment Jupyter Notebook, is more suitable for data exploration than the compiled languages.
\end{itemize}

In order to make graph programming easier, a distributed graph programming framework should satisfy three usability criteria: 1) provide a cross-platform unified programming interface,  2) make distributed computing details transparent to users, and 3) support Python as the programming language.

\begin{table*}
		\centering
		\caption{Comparison of Distributed Graph Processing Systems/Frameworks}
		\label{tab:comparison_distributed_graph_processing_systems_and_frameworks}
		\footnotesize
		\begin{tabular}{x{8em}x{8em}x{8em}x{6em}x{8em}x{8em}x{8em}}\toprule
			System/Framework & Programming Model & Underlying Platform & Programming Language & Distributed Transparency & Interactive Execution & Development Environment \\ \midrule
			Giraph \cite{giraph} & Pregel & Hadoop & Java & $\times$ & $\times$ &IDE \\
			GraphX \cite{graphx} & GAS & Spark & Scala & $\times$ & $\checkmark$                    & IDE + Notebook \\
			Gemini \cite{gemini} & Push-Pull & MPI & C++ & $\times$ & $\times$ & IDE \\
			PowerGraph \cite{powergraph} & GAS & MPI & C++ & $\times$ & $\times$ & IDE \\
			PowerLyra \cite{powerlyra} & GAS & MPI & C++ & $\times$ & $\times$ & IDE \\
			KDT \cite{KDT} & Linear Algebra & MPI & Python & $\checkmark$ & $\checkmark$ & IDE + Notebook \\
			TinkerPop \cite{tinkerpop} & Pregel & Multiple & Java & $\checkmark$ & $\times$ & IDE \\\midrule
			\emph{UniGPS} & VCProg & Multiple & Python & $\checkmark$ & $\checkmark$ & IDE + Notebook \\ \bottomrule
		\end{tabular}

\end{table*}

Unfortunately, as shown in Table~\ref{tab:comparison_distributed_graph_processing_systems_and_frameworks}, the mainstream distributed graph processing systems hardly meet the three criteria simultaneously.
Although KDT \cite{KDT} satisfies the three criteria, the expression power of its linear algebraic programming model is limited, only supporting user-defined functions in several semiring methods like SpGEMM.
TinkerPop \cite{tinkerpop} is a cross-platform graph programming framework, but it only supports Pregel as the programming model for \emph{graph processing}.
It cannot integrate with systems that adopt other programming models (like GraphX and Gemini).
Furthermore, it only supports Java for graph processing.

\subsection{Contributions}

To enhance the usability of distributed graph processing, we propose a cross-platform unified distributed graph programming model \emph{VCProg}.
We summarize the common features of the typical graph programming models Pregel, GAS, and Push-Pull and further propose the VCProg programming model to unify them.
VCProg is vertex-centric.
It regards graph processing as an iterative update of vertex properties.
Each iteration consists of three phases: merging messages, updating vertex properties, and sending messages.
VCProg is compatible with Pregel, GAS, and Push-Pull models.
Programs written with VCProg can be executed by the compatible distributed graph processing systems (like Giraph, GraphX, and Gemini) without modification, achieving the goal of ``Write Once, Run Anywhere.''

Based on the unified programming model VCProg, we further design a unified distributed graph programming framework \emph{UniGPS} that satisfies the three usability criteria.
\begin{enumerate}
	\item UniGPS provides users with unified programming interfaces. Programs written with UniGPS can be executed by the mainstream distributed graph processing systems Giraph, GraphX, and Gemini without modification.

	\item The programming interfaces are platform-independent, hiding details of the underlying distributed computing from users.

	\item UniGPS adopts Python as the programming language. To enable the Java/C++-based distributed graph processing systems to call user-defined methods written in Python, we propose an execution isolation mechanism based on interprocess communication. We propose the zero-copy optimization technique based on memory-mapped buffers to reduce data copying overheads during interprocess communication.
\end{enumerate}

We evaluate the usability and performance of UniGPS in a cluster with nine nodes.
UniGPS enables users to conduct distributed graph processing in a user-friendly interactive Python development environment like Jupyter Notebook.
With the help of distributed computing, the graph scale that UniGPS can handle is an order of magnitude higher than that of the serial graph processing library NetworkX \cite{networkx}.
For the same graph, the processing time of UniGPS is much shorter than NetworkX.
UniGPS achieves near-linear data and machine scalability for typical graph algorithms.
% !TEX root=unigps-main.tex
\section{Related Work}

\subsection{Distributed Graph Processing Systems}

A series of distributed graph programming models and processing systems have been proposed to reduce the difficulty of distributed graph processing.
Reference \cite{graph_processing_framework_survey_2018} presents a comprehensive overview of distributed graph programming frameworks.
According to the granularity of parallel computing, the existing graph programming models can be divided into three categories: vertex-centric \cite{vertex_centric_framework_survey}, block-centric \cite{blogel}, and subgraph-centric \cite{nscale} \cite{gthinker}.
The vertex-centric models are the most thoroughly studied, typical models including Pregel \cite{pregel}, GAS \cite{powergraph}, and Push-Pull \cite{gemini}.
However, even the same programming model has different APIs in different systems, lacking the cross-platform feature.

TinkerPop \cite{tinkerpop} is a graph programming framework that integrates multiple graph databases and processing systems with unified programming interfaces.
TinkerPop proposes Gremlin \cite{gremlin} as its platform-independent query language.
However, TinkerPop's graph processing framework GraphComputer only supports Java as the programming language and adopts Pregel as the graph programming model, not compatible with GAS, Push-Pull, and other programming models.
It can only integrate the Pregel-based graph processing systems.

\subsection{Graph Processing Libraries for Python}

Python is popular among data analysts \cite{top_programming_language_survey_ieee} \cite{top_programming_language_survey_kdnuggets} due to its high usability and development efficiency.
Many Python graph processing libraries have been developed for the single-machine environment \cite{python_packages_for_graph_analysis}.
NetworkX \cite{networkx} and graph-tool \cite{graphtool} are two popular libraries \cite{python_library_popularity_survey}.
However, their performance is limited by the computing power and memory capacity of a single machine.
They can hardly process large-scale graphs.

A possible way to process big graphs in Python is to use general-purpose distributed computing systems like Dask and PySpark.
Due to the lack of specialized encapsulation and optimization for graph processing, users have to manually manage graph data and message exchange between vertices, increasing the programming difficulty.

% !TeX spellcheck = en_US
% !TEX root = unigps-main.tex
\section{Unified Graph Programming Model}

Among the existing distributed graph programming models, the vertex-centric models (like Pregel \cite{pregel} and GAS \cite{powergraph}) are the most thoroughly studied \cite{graph_processing_framework_survey_2018}.
However, different programming models have different APIs, increasing the learning burden and bringing extra programming migration costs.
In order to overcome the drawback, we summarize the common features of three typical vertex-centric graph programming models Pregel \cite{pregel}, GAS \cite{powergraph}, and Push-Pull \cite{gemini}.
We design a vertex-centric unified graph programming model \emph{VCProg} based on the common features.

\subsection{Common Features of Vertex-Centric Programming Models}

Pregel, GAS, and Push-Pull have two salient common features: iterative calculation and three-phase update.
The programming models express graph processing as an iterative update process of vertex properties.
Vertices exchange data by sending messages between iterations.

The whole process consists of several rounds of iterations.
In each iteration, every vertex updates its property based on the properties of itself, its neighbors, and the received messages.
The iterations continue until convergence.
There are two popular convergence conditions: all vertices are inactive, or the iteration reaches the maximum number of rounds.

\begin{figure}
    \centering
	\includegraphics[width=0.8\columnwidth]{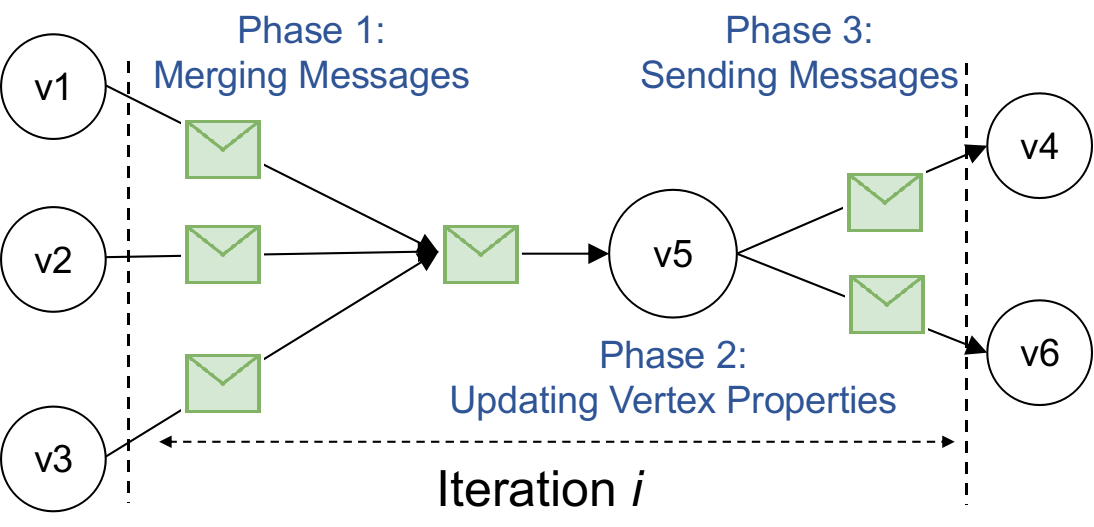}
	\caption{Three phases of an iteration in vertex-centric programming models.}
	\label{fig:three_phases_of_a_vertex_centric_iteration}
\end{figure}

In each iteration, the process of every vertex consists of three phases: merging messages, updating vertex properties, and sending messages, as shown in Fig.~\ref{fig:three_phases_of_a_vertex_centric_iteration}.
Firstly, each vertex receives messages from its incoming neighbors and merges the messages into a single message.
Secondly, each vertex updates its property based on the merged message and its current property.
Finally, each vertex sends messages to its outgoing neighbors based on its updated property and the properties of its adjacent outgoing edges.

Inspired by the features, we propose a vertex-centric unified graph programming model \emph{VCProg} that is compatible with Pregel, GAS, and Push-Pull at the same time.

\subsection{Data Model of VCProg}

VCProg adopts the property graph as its data model.
Each vertex (edge) has an attached property that is a record containing several fields.
All vertex (edge) properties have the same schema.
Messages exchanged between vertices are also records.
All messages have the same schema.

Before iterations begin, each vertex (edge) initializes its property based on the input data.
During iterations, the vertex properties are updated while the edge properties remain unchanged.
After iterations, the vertex properties store the processing results.
The vertex properties are output to files in a tabular form.

\subsection{Application Programming Interface of VCProg}

\begin{figure}
	\includegraphics[width=0.8\columnwidth]{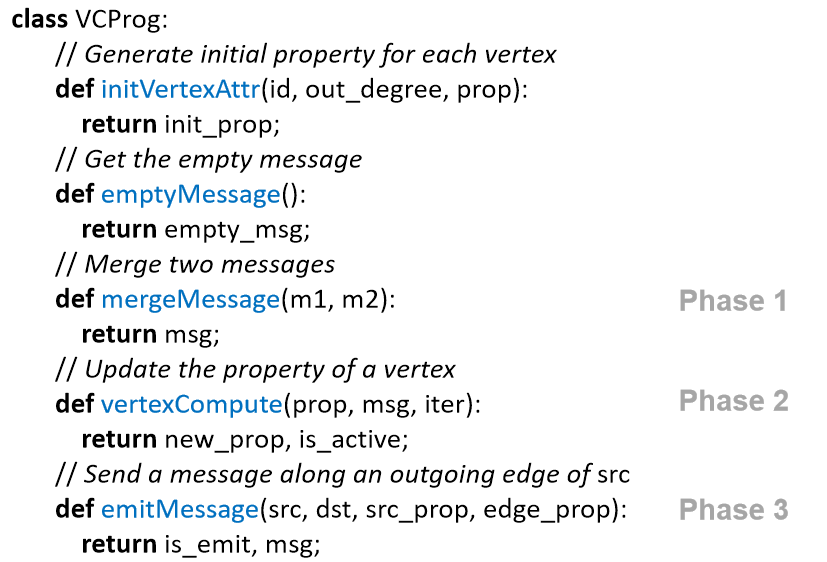}
	\caption{Application programming interface of VCProg in Python.}
	\label{fig:api_of_vcprog}
\end{figure}

VCProg provides its Python API in the form of an abstract base class \texttt{VCProg} as shown in Fig.~\ref{fig:api_of_vcprog}.
All methods of the base class are abstract.
To write distributed graph processing programs, users need to inherit the base class and implement the abstract methods according to algorithmic logic.
Users define the behavior of the three phases in each iteration with the \texttt{mergeMessage}, \texttt{vertexCompute}, and \texttt{emitMessage} method, respectively.

The \texttt{mergeMessage} method combines two message records \texttt{m1} and \texttt{m2} into a single message record \texttt{msg}.
The message order should be interchangeable: \texttt{mergeMessage(m1, m2)} = \texttt{mergeMessage(m2, m1)}.

The \texttt{vertexCompute} method generates the updated vertex property for a vertex in each iteration.
It receives the vertex property \texttt{prop} from the previous iteration, the merged message \texttt{msg}, and the current iteration number \texttt{iter} as the parameters.
The method returns the updated vertex property \texttt{new\_attr} and a flag \texttt{is\_active} to indicate whether the vertex will be active in the next iteration.

The \texttt{emitMessage} method determines whether to send a message \texttt{is\_emit} and the content of the message \texttt{msg} for the edge (\texttt{src}, \texttt{dst}), based on the source vertex's property \texttt{src\_prop} and the edge's property \texttt{edge\_prop}.

VCProg further uses the \texttt{initVertexAttr} method to initialize the property of each vertex before iterations, based on the vertex ID \texttt{id}, the property in the input \texttt{prop} and the outgoing degree \texttt{out\_degree}.
The \texttt{emptyMessage} method returns a global read-only empty message record \texttt{empty\_msg}.
The empty message is a special message that is idempotent for merging.
For any message \texttt{m}, it should satisfy \texttt{mergeMessage(m, empty\_msg)} = \texttt{m}.

\begin{figure}
	\includegraphics[width=0.8\columnwidth]{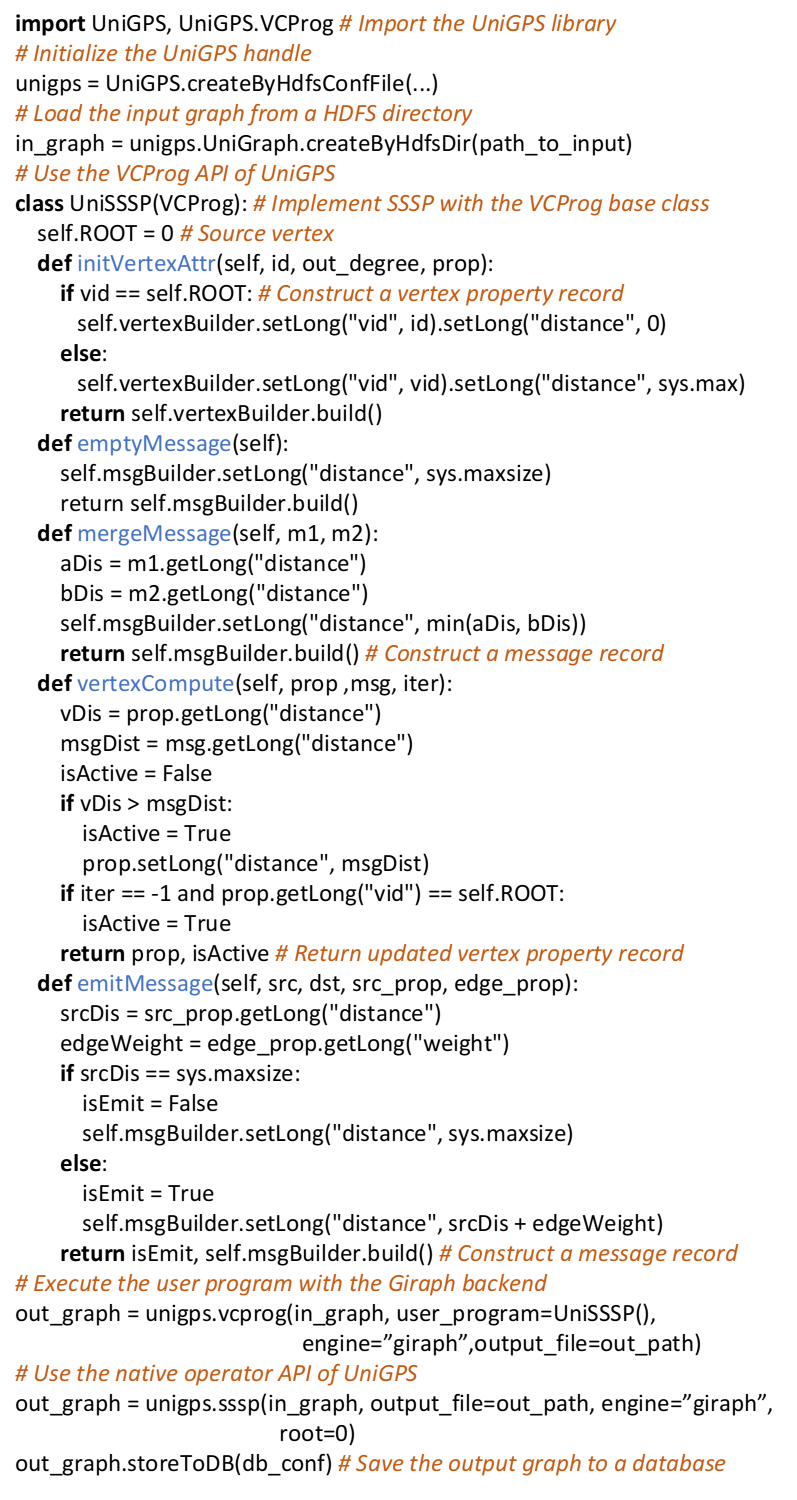}
	\caption{Demo program of UniGPS to calculate the single-source shortest path (SSSP) in Python.}
	\label{fig:toy_example_of_unigps}
\end{figure}

Fig.~\ref{fig:toy_example_of_unigps} shows a demo program that uses VCProg to implement the Bellman-Ford single-source shortest path calculation (UniSSSP).
The API of VCProg is easy to use and independent from distributed graph processing systems, ensuring the transparency of distributed processing to users.

\subsection{Execution Semantics of VCProg}

Algorithm~\ref{alg:workflow_of_vcprog} shows the workflow of VCProg.
VCProg receives a graph $G=(V, E)$ as the input, where $V$ and $E$ are the vertex and edge set of $G$, respectively.
Each vertex $v$ (edge $e$) in $G$ has an attached property $v.value$ ($e.value$).
The user needs to specify the maximum number of iterations \emph{MAX\_ITER} as the hyper-parameter.
The user provides its program in the form of an instance object \emph{VP} of the VCProg base class.
\emph{VP} implements all the abstract methods.
\emph{VP} is read-only and shared by all vertices.

VCProg iterates for at most \emph{MAX\_ITER} rounds (line 10).
In each round, every vertex $v$ will be either active or inactive.
$v$ is active if and only if $v$ is set as active in the previous round or $v$ receives any message.
Only active vertices participate in the current iteration (Line 7 to 16).
The \texttt{vertexCompute} method determines whether the vertex remains active in the next round.
If all vertices are inactive in a round (Line 17), the iteration converges early.
VCProg triggers a global barrier implicitly at the end of each round.

\begin{algorithm}
	\caption{Workflow of the VCProg Programming Model}
	\label{alg:workflow_of_vcprog}
	\footnotesize
	\textbf{Input}: $G=(V,E)$, \emph{MAX\_ITER}, \emph{VP};	\textbf{Output}: $G$.
	\begin{algorithmic}[1]
		\State $empty\_msg \leftarrow$ \textit{VP.emptyMessage}(); \Comment{Global empty message}
		\ForAll{$v \in V$} \textbf{in parallel} \Comment{Initialize vertex properties}
			\State $v.value$ $\gets $\textit{VP.initVertexAttr}($v.\textit{ID}$, len($v.out\_edges$), $v.value$);
		\EndFor
		\For{$iter \leftarrow$ 1 to \emph{MAX\_ITER}}
			\State $num\_active \leftarrow 0$; \Comment{Number of active vertices}
			\ForAll{$v \in V$ that $v$ is active or $v$ receives any message} \textbf{in parallel}
				\State $msg \gets empty\_msg$;
				\For{every message $m$ received by $v$}
					\State $msg \gets $ \textit{VP.mergeMsg}($msg$, $m$);
				\EndFor
				\State $v.value$, $v.is\_active$ $\gets$ \textit{VP.vertexCompute}($v.value$, $msg$, $iter$);
				\If{$v.is\_active$ = TRUE} \Comment{Only for active vertices}
					\State $num\_active \gets num\_active +1$;
					\ForAll{$e \in v.out\_edges$} \Comment{For outgoing adjacent edges}
						\State $is\_emit, msg \gets$ \textit{VP.emitMessage}($v$.\textit{ID}, $e$.\textit{target\_ID}, $v.value$, $e.value$);
						\If{$is\_emit$ = TRUE}
							\State \Call{SendMessage}{$e$.\emph{target\_ID}, $msg$}
						\EndIf
					\EndFor
				\EndIf
			\EndFor
			\If{$num\_active = 0$}
				\State \textbf{break;} \Comment{The iteration converges and terminates early}
			\EndIf
		\EndFor
	\end{algorithmic}
\end{algorithm}

\subsection{Compatibility with Other Programming Models}

\begin{figure}
	\subfloat[Pregel]{\includegraphics[width=\columnwidth]{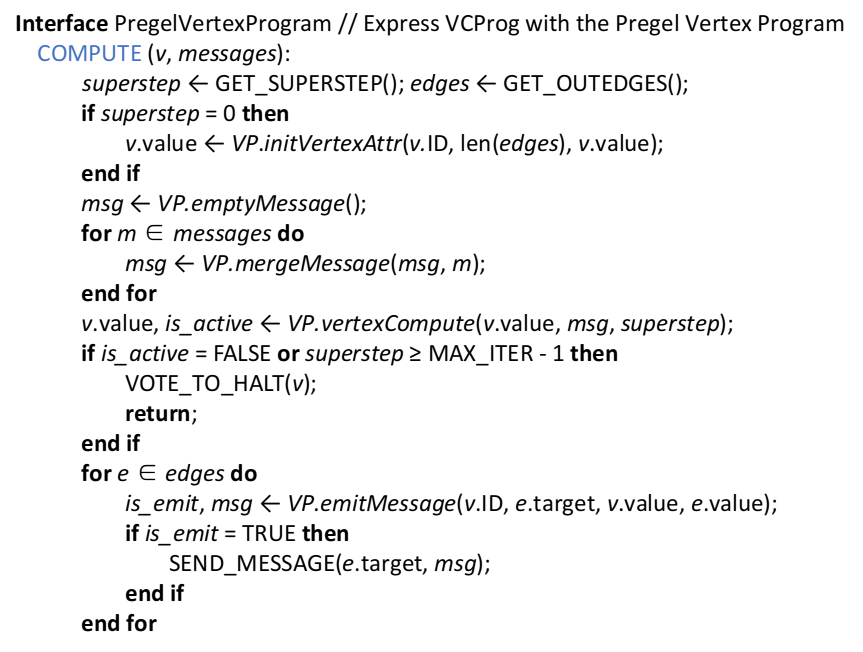}}\\
	\subfloat[GAS\label{fig:express_vcprog_in_gas}]{\includegraphics[width=\columnwidth]{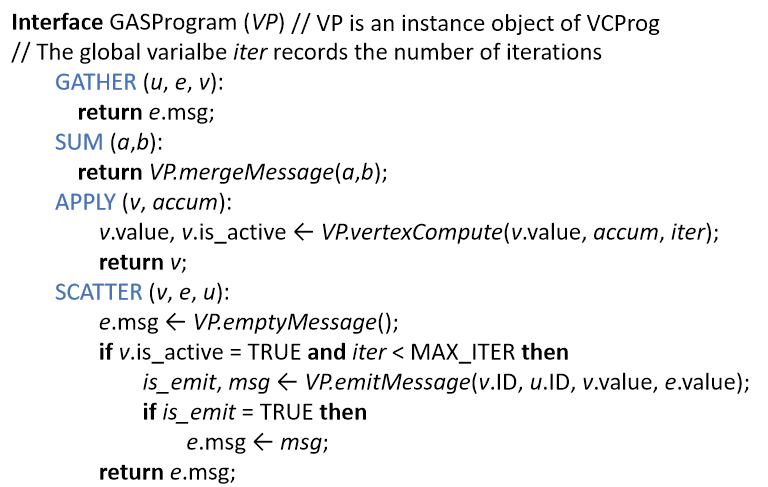}}\\
	\subfloat[Push-Pull (Dense Mode)]{\includegraphics[width=\columnwidth]{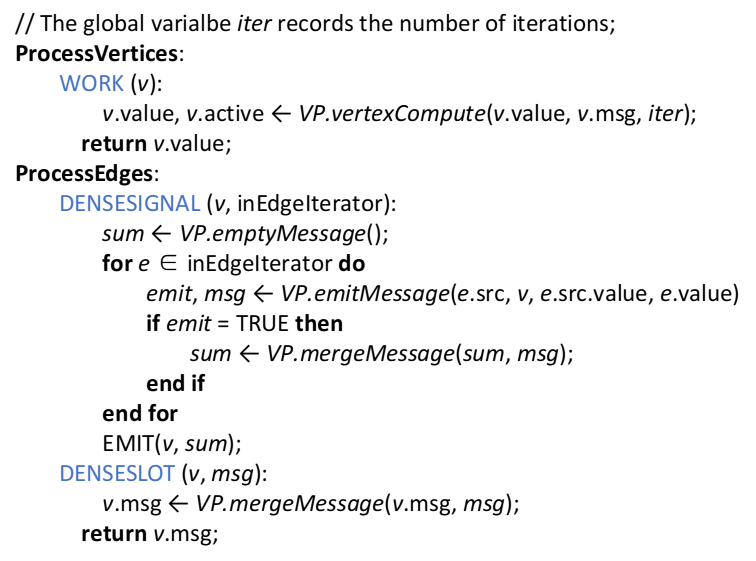}}
	\caption{Express the workflow of the VCProg programming model (\emph{VP}) with the typical vertex-centric graph programming models.}
	\label{fig:express_vcprog}
\end{figure}

VCProg is compatible with the typical vertex-centric graph programming models Pregel
\cite{pregel}, GAS \cite{powergraph} and Push-Pull \cite{gemini}.
The workflow of VCProg can be equivalently expressed by these models.
Fig.~\ref{fig:express_vcprog} shows how to use Pregel, GAS, and Push-Pull to achieve the same execution semantics of VProg given a user-defined VCProg instance object \emph{VP}.
According to the conversion, the corresponding distributed graph processing system can execute a VCProg program \emph{VP} to get the desired output.
By this way, VCProg provides the capability of cross-platform execution.

% !TeX spellcheck = en_US
% !TEX root = unigps-main.tex

\section{Unified Graph Programming Framework}

Based on the unified graph programming model VCProg, we design and implement a unified graph programming framework \emph{UniGPS}.
Users can use UniGPS in \emph{interactive} Python development environments like Jupyter Notebook.
To further support cross-platform program execution, we propose an interprocess-communication-based mechanism to isolate the execution of user programs (written in Python) from the underlying distributed graph processing systems.

\subsection{System Architecture}

\begin{figure}
	\includegraphics[width=\columnwidth]{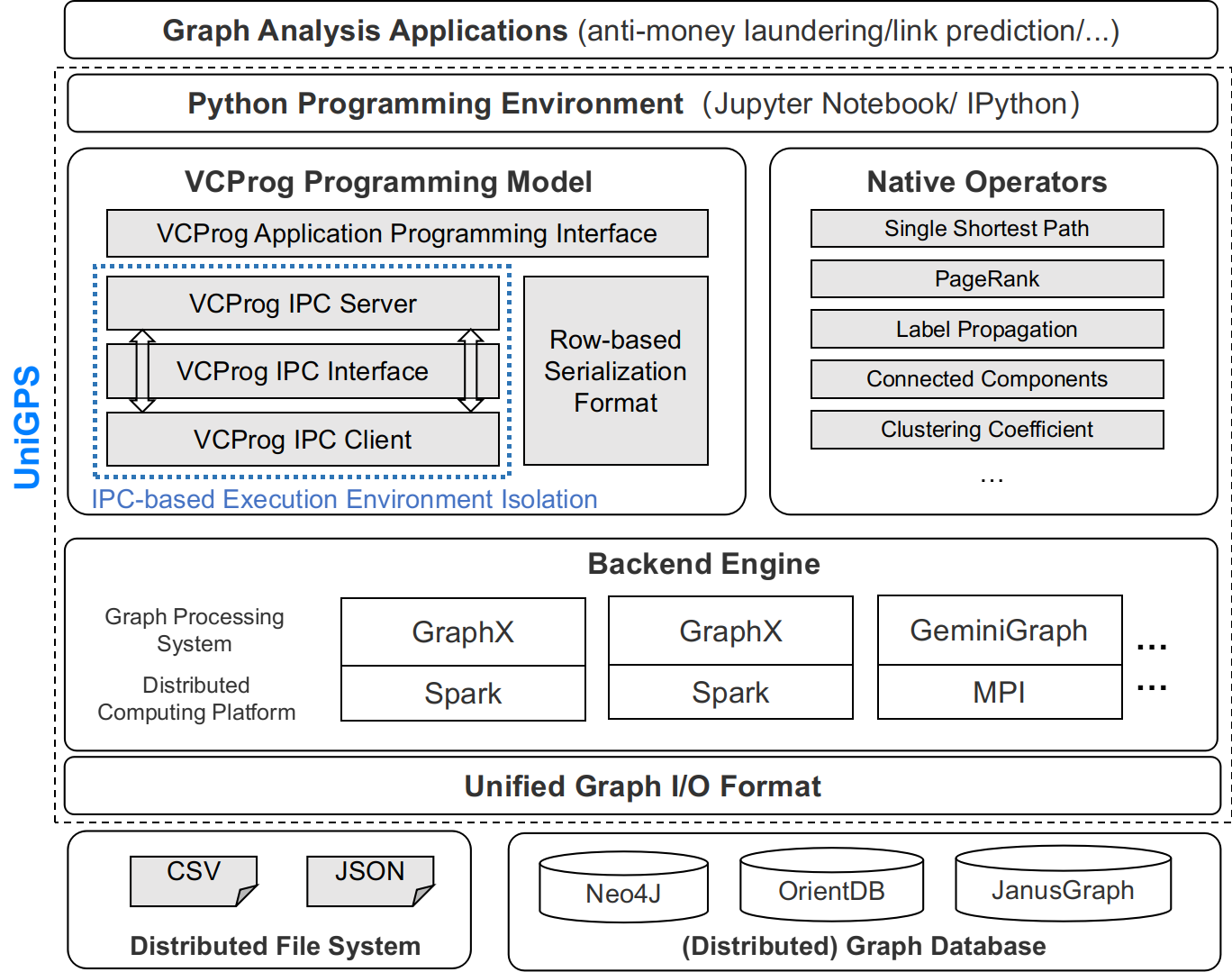}
	\caption{System architecture of UniGPS.}
	\label{fig:system_architecture_of_unigps}
\end{figure}

Fig.~\ref{fig:system_architecture_of_unigps} shows the system architecture of UniGPS.
UniGPS consists of four modules: VCProg programming model, native operators, backend engine, and the unified graph I/O format.

The VCProg programming model module provides the VCProg API and runs custom user programs.
Considering Java/C++-based distributed graph processing systems cannot execute Python-based user programs directly, we propose an interprocess-communication-based mechanism to isolate the execution environment of the two parts.
We adopt a row-based serialization format to serialize property/message records during communication.
We will elaborate on the mechanism later in Section~\ref{sec:execution_environment_isolation}.

The native operator module contains some frequently-used pre-compiled graph operators.
UniGPS natively implements every operator for every distributed graph processing system in advance.
UniGPS provides a unified platform-independent API for each operator.
Every API contains an \texttt{engine} parameter to select the preferred graph processing system.

The backend engine module integrates VCProg-compatible distributed graph processing systems into UniGPS.
Currently, our prototype of UniGPS supports Giraph, GraphX, and Gemini.
UniGPS can integrate with other systems that are compatible with the VCProg programming model.
Each system is regarded as a backend engine to run VCProg programs and native operators.

The unified graph I/O format module uses a unified graph serialization format (like GraphSON
\cite{tinkerpop}) to decouple external data sources and distributed graph processing systems.
If we want to support $M$ systems reading/writing $N$ data sources, we have to implement $M*N$ I/O formats without the unified format.
Using the unified format as an intermediate format, we only need to implement $M+N$ I/O formats, significantly reducing the development and maintenance costs of UniGPS.

\subsection{Application Programming Interface of UniGPS}

UniGPS appears as a library in Python.
Fig.~\ref{fig:toy_example_of_unigps} demonstrates how to conduct the single-source shortest path calculation in UniGPS.
Users need to import the UniGPS library first and initialize a handle of UniGPS \texttt{unigps} with the configuration file.
UniGPS supports loading/saving graphs from/to external data sources like HDFS and graph databases.

UniGPS provides two kinds of APIs to conduct graph processing: VCProg APIs and native operator APIs.
The VCProg APIs allow users to develop customized processing programs.
The native operator APIs allow users to call the built-in native implementation of frequently-used processing operators like PageRank and single-source shortest path (SSSP).
All UniGPS's APIs are platform-independent.
Users can execute UniGPS programs with different distributed graph processing systems by just changing the engine parameter in each API.

\subsection{Interprocess Communication Based Execution Environment Isolation}
\label{sec:execution_environment_isolation}

The distributed graph processing systems written in Java, Scala, or C++ call the user-defined functions via function/method invocation provided by the programming language itself.
It requires users to develop their programs in the same programming language as the systems.

To enable the existing Java/Scala/C++-based systems to call the user-defined methods of VCProg written in Python, we propose an execution environment isolation mechanism based on interprocess communication (IPC).
As shown in Fig.~\ref{fig:system_architecture_of_unigps}, the mechanism contains three modules: IPC server, IPC interface, and IPC client.
The IPC server is a Python process that contains the user-given VCProg instance object.
The IPC client embeds itself in the worker processes of distributed graph processing systems.
The IPC client and server communicate through the IPC interface.
The IPC interface allows the IPC client to remotely call the methods of the VCProg object in the IPC server.

\subsubsection{Workflow of VCProg Jobs}

\begin{figure}
    \centering
    \includegraphics[width=0.85\columnwidth]{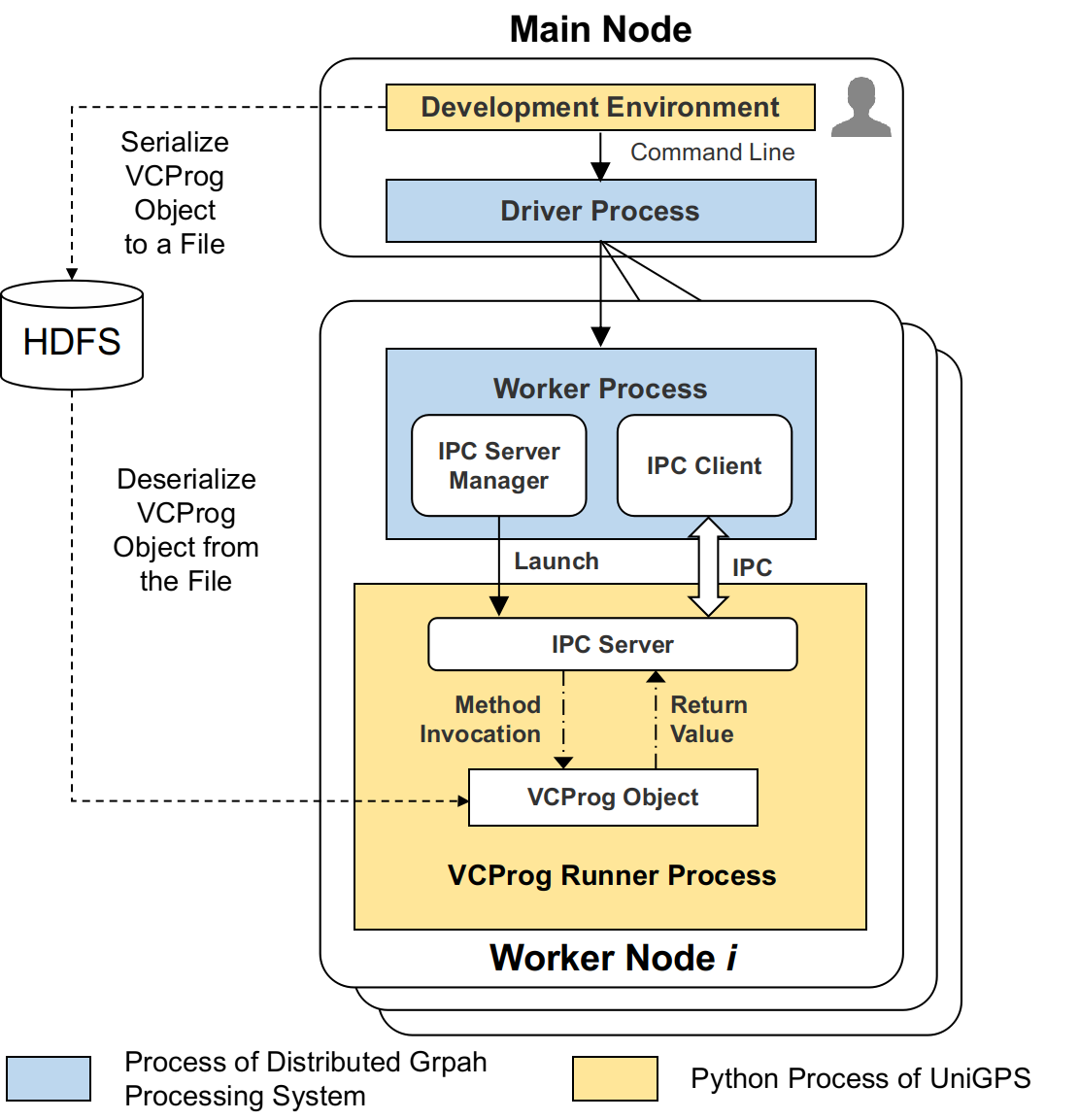}
    \caption{Execution environment isolation with interprocess communication.}
    \label{fig:execution_environment_isolation}
\end{figure}

With the execution environment isolation mechanism, UniGPS adds several extra steps in the workflow of a VCProg-based graph processing job, as shown in Fig.~\ref{fig:execution_environment_isolation}.

Before a job starts, UniGPS serializes the VCProg instance object that the user submits to a file, and UniGPS uploads the file to HDFS.

UniGPS starts a distributed graph processing job based on the user-specified backend engine specified through the command line.
The job starts its driver process on the main node of the cluster.
The driver process further starts several worker processes on the worker nodes of the cluster.
Taking GraphX as an example, the Spark driver process starts Spark executors through the cluster resource manager YARN.

Every worker process launches a dual VCProg runner process.
The VCProg runner process deserializes the VCProg object from the file stored on HDFS and creates an IPC server based on the VCProg object.
The worker process creates an IPC client and establishes the connection with the IPC server.

After the initialization steps above accomplish, the driver process starts running the distributed graph processing job according to the workflow of the backend engine system.
For example, the GAS-model-based systems follow the workflow in Fig.~\ref{fig:express_vcprog_in_gas} to run the job.
During the job execution, when a worker process needs to call a method of the VCProg instance object, it triggers a remote procedure call through the IPC client to the VCProg object in the IPC server.

The execution isolation mechanism enables UniGPS to support Python for any distributed graph processing system.
It hides the details of distributed processing from users and supports cross-platform execution of user programs.

\subsubsection{Interprocess Communication Optimization}

The performance of the execution isolation mechanism significantly affects the efficiency of UniGPS.
We can implement the mechanism with any general-purpose remote procedure call (RPC) framework.
However, the popular RPC frameworks like gRPC are based on the network stack.
They trigger system calls and copy data buffers between the user and kernel space multiple times during a remote method invocation.

\begin{figure}
    \centering
    \includegraphics[width=0.8\columnwidth]{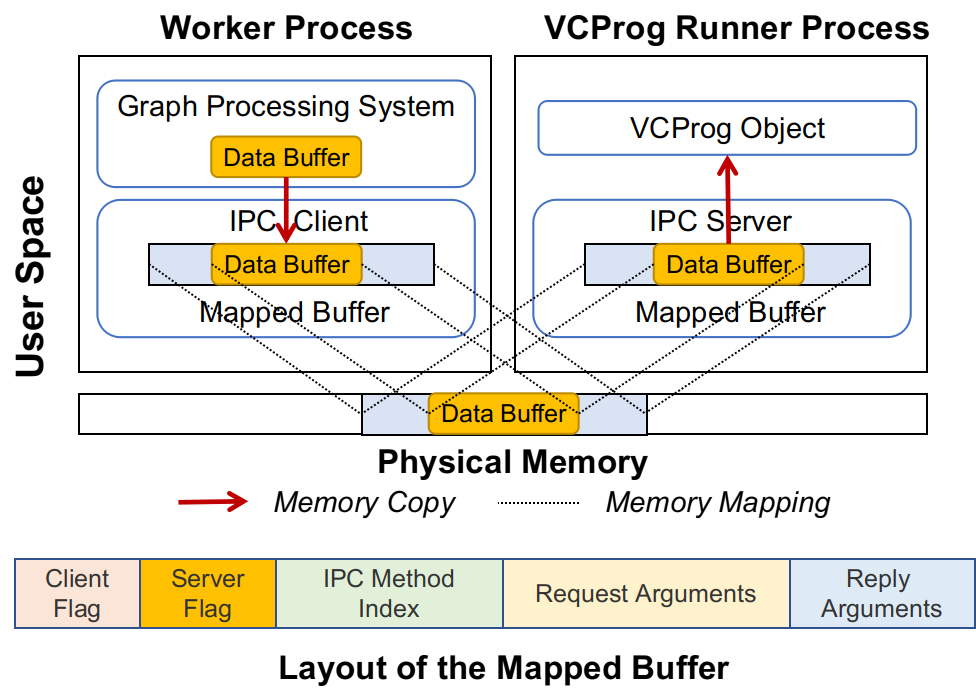}
    \caption{Memory-mapping-based interprocess communication.}
    \label{fig_zero_copy_ipc_implementation}
\end{figure}

In order to reduce the overhead of data copies, we adopt a zero-copy IPC technique to implement the execution environment isolation mechanism.
UniGPS creates a mapped buffer on both sides of the IPC client and server, as shown in Fig.~\ref{fig_zero_copy_ipc_implementation}.
We map the two buffers to the same region in the physical memory through the memory mapping file mechanism provided by Linux.
The reading/writing of the mapped buffers directly operates on the corresponding region in the physical memory.
Changes to one of the buffers immediately reflect in the other buffer without any data copy.
In other words, the communication between the IPC client and server does not involve any data copy (i.e., \emph{zero copy}), and the communication occurs in the user space, avoiding the overheads of kernel-space switching.
With the mapped buffers, the problem of implementing RPC becomes how to organize concurrent reading/writing of the mapped buffers on both sides.

UniGPS uses the memory layout in Fig.~\ref{fig_zero_copy_ipc_implementation} to manage the mapped buffers.
The client flag indicates whether the client is ready to prepare the IPC method index and the request arguments.
The server flag indicates whether the server finishes the method invocation.
The IPC method index indicates which method of the VCProg object the IPC client calls.

Since RPC invocations usually finish quickly in VCProg, we use the busy waiting mechanism to synchronize between the IPC client and server.
When the client starts an RPC invocation, it first sets the client flag and then repeatedly checks whether the server flag becomes ready.
The server side also repeatedly checks the client flag.
Once the client flag becomes ready, the server processes the RPC immediately.
When the server finishes, it sets the server flag to notify the client.
Compared with the lock-based mechanism, the busy waiting avoids triggering system calls.
UniGPS uses the thread yield mechanism in busy waiting to actively give up invalid CPU time slices, reducing the waste in CPU cycles.
% !TEX root = unigps-main.tex
% !TEX spellcheck=en_US
\section{Performance Evaluation}

The section evaluates the usability and scalability of UniGPS and compares its computational performance with the stand-alone Python graph processing library NetworkX.

\subsection{Experimental Environment}

All experiments were conducted in a cluster with nine nodes (1 main + 8 workers) connected via 1Gbps ethernet.
Each node was equipped with eight physical cores, 40 GB memory, and 1.8 TB HDD hard disk.
We launched a Ubuntu 20.04 container for each node to provide an isolated experimental environment.
We used CPython 3.7.4 as the Python interpreter.

UniGPS integrated three distributed graph processing systems as the backend engines: Giraph \cite{giraph} (version 1.3.0 with Hadoop 2.7.7 and OpenJDK 1.8), GraphX \cite{graphx} (with Spark 2.1.2 and Scala 2.11.8), and Gemini \cite{gemini} (compiled with GCC 9.3.0).
For Giraph, UniGPS launched eight workers for each node and allocated 4 GB memory for each worker.
For GraphX, UniGPS launched eight Spark executors for each node and allocated 4 GB memory for each executor.
For Gemini, each node ran eight single-thread worker processes.

\begin{figure*}
    \centering
    \subfloat[\label{fig_unigps_perf_comparison}]{\includegraphics[width=0.8\textwidth]{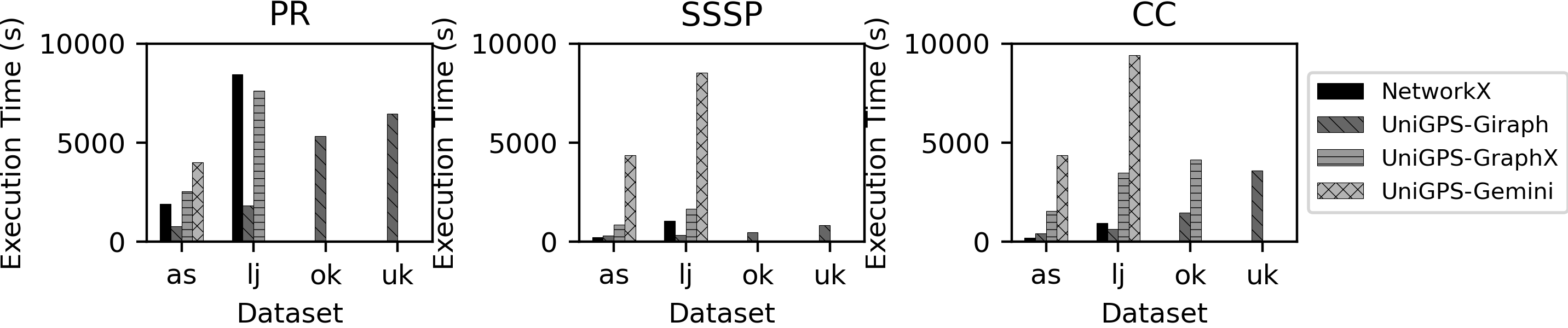}}\\
    \subfloat[\label{fig_data_scalability}]{\includegraphics[width=0.32\textwidth]{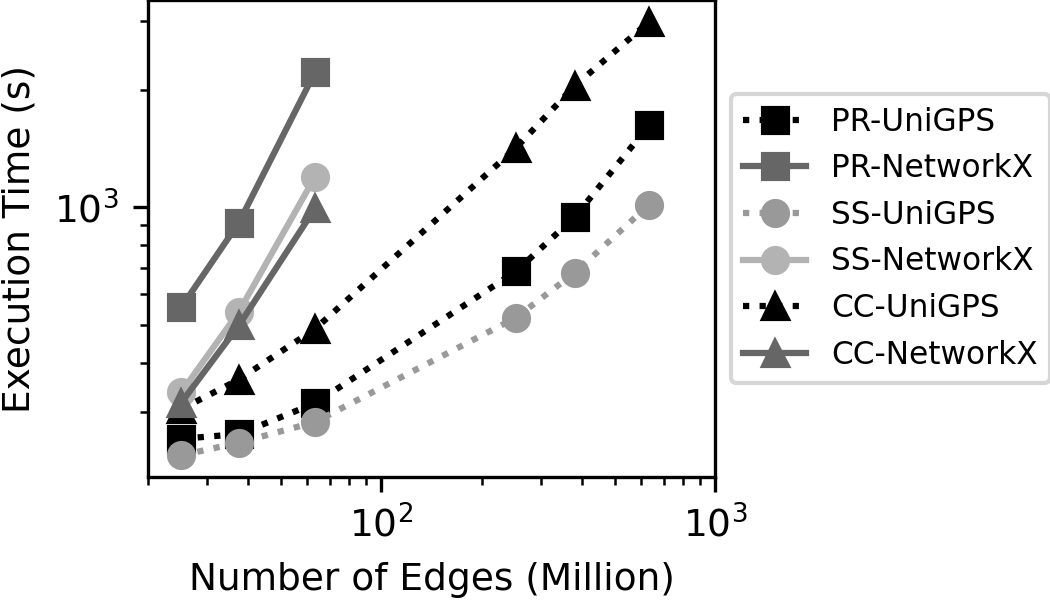}}\hfill
    \subfloat[\label{fig_machine_scalability}]{\includegraphics[width=0.28\textwidth]{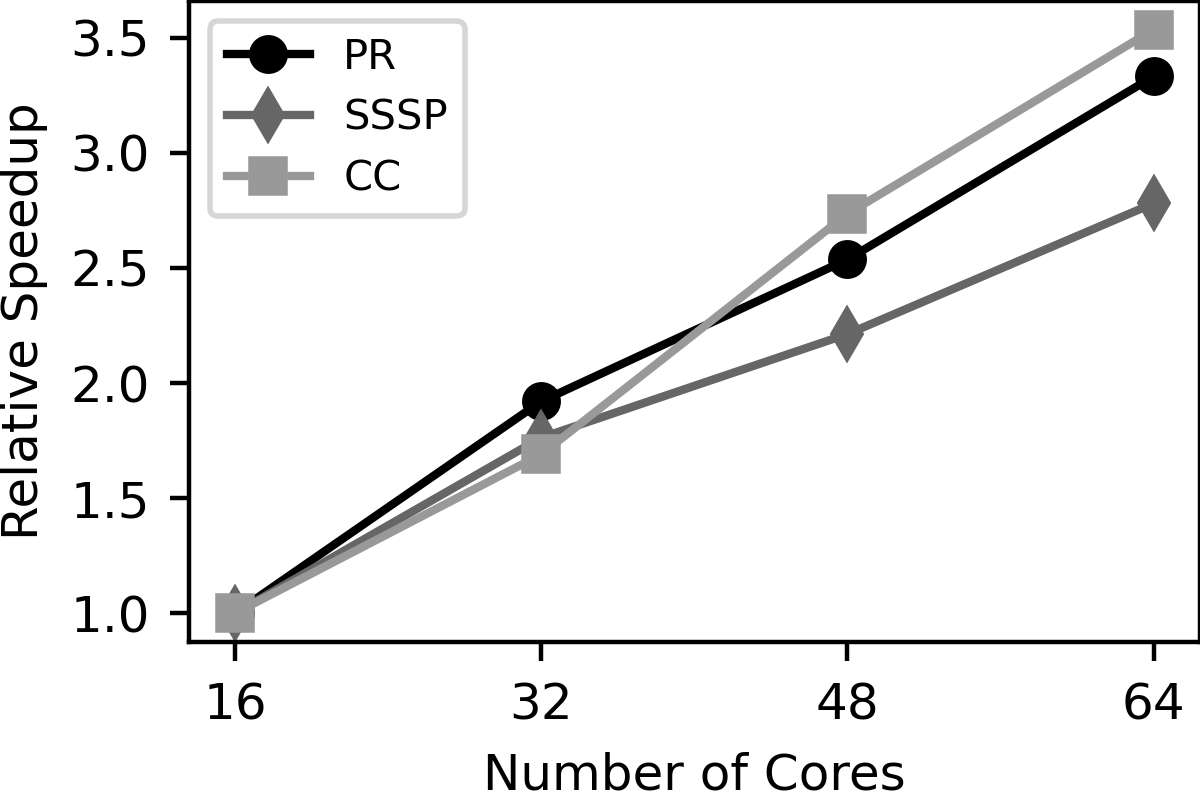}}\hfill
    \subfloat[\label{fig_ipc_perf_comparison}]{\includegraphics[width=0.32\textwidth]{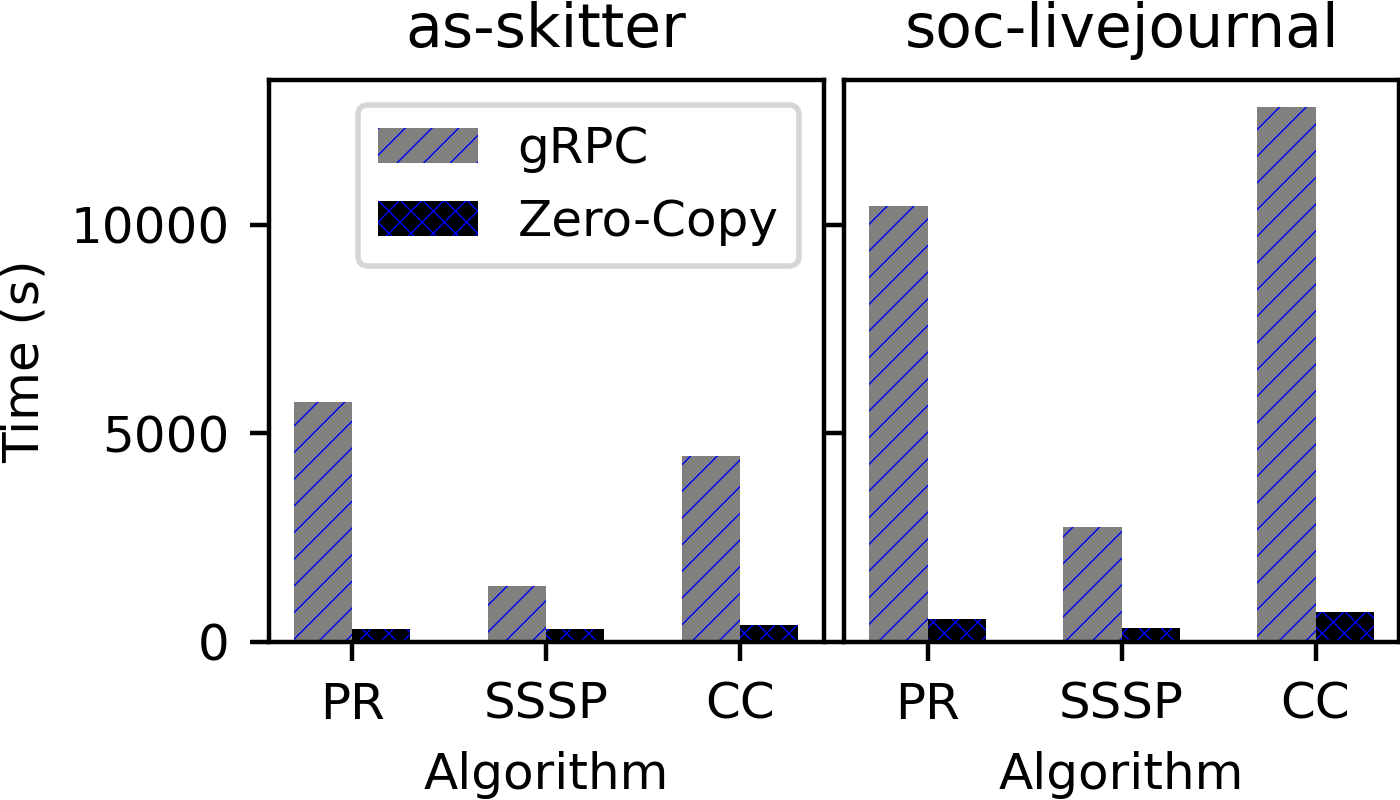}}
    \caption{Performance evaluation of UniGPS on typical graph algorithms: PageRank (PR), single-source shortest path (SSSP), and connected components (CC). (a) Performance comparison between UniGPS with different engines and NetworkX. (b) Data scalability of UniGPS and NetworkX. (c) Machine scalability of UniGPS. (d) Effects of IPC optimization.}
\end{figure*}

\subsection{Usability Comparison}
Table~\ref{tab:comparison_distributed_graph_processing_systems_and_frameworks} compares the usability of UniGPS and the existing distributed graph processing systems/frameworks.
UniGPS supports the Python language and can be used in an \emph{interactive} development environment like Jupyter Notebook, improving the productivity of program development and debugging.
The APIs of UniGPS are platform-independent, hiding details of distributed computing from users.
UniGPS and KDT are the only two frameworks that satisfy the three usability criteria discussed in Section~\ref{sec_backgroud}.
However, the customizability of KDT is more limited than UniGPS.
KDT only supports using user-defined functions for several semiring methods.
UniGPS can customize the methods of all three phases in each iteration.

\subsection{Performance Comparison}

We compared the execution time of UniGPS with the serial Python graph processing library NetworkX \cite{networkx} (version 2.5) on four real-world graph datasets (as shown in Table~\ref{tab_overview_realworld_graph}).
Fig.~\ref{fig_unigps_perf_comparison} shows the experimental results on several typical graph algorithms.
For UniGPS, we implemented the typical algorithms with the VCProg API.
For NetworkX, we call the built-in algorithm operators.
If the execution time exceeded three hours (i.e., timeout) or the program crashed, the test case was not shown.

The experimental results indicated that UniGPS could handle much larger datasets than NetworkX.
NetworkX could not process the large datasets \texttt{ok} and \texttt{uk} due to the out-of-memory exception.
UniGPS with the Giraph engine could handle all datasets with the help of distributed processing.
On the \texttt{lj} dataset, the execution time of UniGPS with the Giraph engine was 21.76\% (PageRank), 31.07\% (SSSP), and 70.23\% (connected components) of NetworkX.
UniGPS met timeout with the GraphX and Gemini engines.
The edge-parallel design of GraphX and Gemini made IPC overheads more obvious.
UniGPS was more suitable to work with Giraph.

\begin{table}
    \caption{Overview of Real-world Graph Datasets}
    \label{tab_overview_realworld_graph}
    \scriptsize
    \centering
    \begin{tabular}{@{}ccccc@{}}
        \toprule
        Dataset                       & $|V|$   & $|E|$    & Directed       & Source           \\ \midrule
        as-skitter (\texttt{as}) \cite{snapnets}      & 1.7M  & 22.2M  & No & Computer Network \\
        soc-livejournal (\texttt{lj}) \cite{snapnets} & 4.8M  & 69.0M  & Yes   & Social Network   \\
        com-orkut (\texttt{ok}) \cite{snapnets}       & 3.1M  & 234.4M & No & Social Network   \\
        uk-2002 (\texttt{uk}) \cite{websoft_lab}         & 18.5M & 298.1M & Yes   & WWW              \\ \bottomrule
    \end{tabular}
\end{table}

\subsection{Data Scalability}

To evaluate the data scalability of UniGPS, we used the logNormalGraph generator of GraphX to generate random graphs of similar topological characteristics but with different scales.
Fig.~\ref{fig_data_scalability} shows the execution time of UniGPS (VCProg API with the Giraph engine) and NetworkX on different scales.

The execution time of UniGPS and NetworkX increased near linearly with the number of edges of the graph.
The execution time of UniGPS was much shorter than that of NetworkX.
The advantages became more obvious as the graph scale increased.
UniGPS showed near-linear data scalability while NetworkX crashed due to the out-of-memory exception on big graphs.
The graph scale that UniGPS could handle was an order of magnitude higher than NetworkX.
UniGPS could help users process big graphs far beyond the memory capacity of a single machine with high usability.

\subsection{Machine Scalability}

We measured the execution time of UniGPS (VCProg API with the Giraph engine) with different numbers of CPU cores.
Fig.~\ref{fig_machine_scalability} converts the execution time into the speedup relative to the case of 16 cores.
The experimental results indicated that UniGPS showed near-linear machine scalability.
The scalability of CC and PR was better than that of SSSP because the two algorithms were more computationally intensive.

\subsection{Effects of IPC Optimization}

Fig.~\ref{fig_ipc_perf_comparison} compares the execution time of UniGPS with two different RPC implementations in the execution environment isolation mechanism: the network-based gRPC and the proposed zero-copy IPC.
The execution time of the zero-copy IPC was much shorter than that of gRPC.
System calls and data copies of gRPC bring high overheads.
Since UniGPS frequently triggers remote procedure calls during execution, adopting the zero-copy IPC can significantly reduce the execution time.

\section{Conclusion and Future Work}

The existing distributed graph processing systems are not friendly enough for data analysts and algorithm engineers.
To increase the usability of distributed graph processing, we proposed a unified graph programming framework UniGPS.
UniGPS provides cross-platform unified programming interfaces, hides distributed computing details from users, and supports Python as the programming language.
UniGPS adopts a vertex-centric unified graph programming model VCProg compatible with the Pregel, GAS, and Push-Pull model.
UniGPS uses an IPC-based execution environment isolation mechanism to enable Java/C++-based systems to call user-defined methods written in Python.

In the future, we plan to propose more techniques to improve the performance of UniGPS.
One possible technique is to organize RPC invocations in a pipeline manner to overlap computing and communication.
High-performance graph I/O is also an interesting topic to investigate.

\section*{Acknowledgment}

This work was supported in part by the National Natural Science Foundation of China (\#U1811461),
the National Key R\&D Program of China (\#2019YFC1711000),
the Open Project of State Key Laboratory for Novel Software Technology (\#KFKT2021B33),
and the Collaborative Innovation Center of Novel Software Technology and Industrialization, Jiangsu, China.
Guanghui Zhu and Yihua Huang are corresponding authors with equal contributions.

\bibliographystyle{IEEEtran}
\bibliography{unigps}

\end{document}